# Screening 2D materials with topological flat bands

Hang Liu[1,2,3], Sheng Meng[2,3,*], and Feng Liu[1,†]

[1] *Department of Materials Science and Engineering, University of Utah, Salt Lake City, Utah 84112, USA*

[2] *Songshan Lake Materials Laboratory, Dongguan, Guangdong 523808, People's Republic of China*

[3] *Beijing National Laboratory for Condensed Matter Physics and Institute of Physics, Chinese Academy of Sciences, Beijing 100190, People's Republic of China*

**Abstract:** Topological flat band (TFB) has been proposed theoretically in various lattice models, to exhibit a rich spectrum of intriguing physical behaviors. However, the experimental demonstration of flat band (FB) properties has been severely hindered by the lack of materials realization. Here, by screening materials from a first-principles materials database, we identify a group of 2D materials with TFBs near the Fermi level, covering some simple line-graph and generalized line-graph FB lattice models. These include the Kagome sublattice of O in $TiO_2$ yielding a spin-unpolarized TFB, and that of V in ferromagnetic $V_3F_8$ yielding a spin-polarized TFB. Monolayer $Nb_3TeCl_7$ and its counterparts from element substitution are found to be breathing-Kagome-lattice crystals. The family of monolayer $III_2VI_3$ compounds exhibit a TFB representing the coloring-triangle lattice model. $ReF_3$, $MnF_3$ and $MnBr_3$ are all predicted to be diatomic-Kagome-lattice crystals, with TFB transitions induced by atomic substitution. Finally, $HgF_2$, $CdF_2$ and $ZnF_2$ are discovered to host dual TFBs in the diamond-octagon lattice. Our findings pave the way to further experimental exploration of eluding FB materials and properties.

*Introduction*. The destructive interference of wavefunctions in a crystal lattice gives rise to a type of electronic band without dispersion, dubbed as topological/singular flat bands (TFBs) [1-5]. Without spin-orbit coupling (SOC), the TFB can be identified by the presence of a band touching point with a dispersive band at a high-symmetric $k$ point, where its singular Bloch wavefunctions in reciprocal space manifest the emergence of topological noncontractible edge states in real space [4, 5]; it differs from an isolated trivial flat band (FB) with nonsingular Bloch wavefunctions, such as the dangling-bond states. With SOC, the degeneracy at the touching point is lifted, leading to the gapped 2D TFB with a nonzero (spin) Chern invariant [4]. The nontrivial topology and the inherently strong electron-electron interaction due to vanishing electron kinetic energy render the TFB a rich spectrum of physical phenomena, such as fractional quantum anomalous Hall effect [6-8], ferromagnetism [9, 10], Wigner crystallization [11, 12], superconductivity [13], excitonic insulator state [14, 15], and excited quantum anomalous/spin Hall effect [16].

Various lattice models have been theoretically proposed to host the TFB. These models are generally based on line-graph construction **[17-19]**, such as the Kagome [6, 19, 20], Lieb [21, 22], breathing-Kagome [23-26], diatomic-Kagome [16, 27], coloring-triangle lattices [28], and the diamond-octagon lattice (i.e., line-graph lattice of Lieb lattice) [2, 29]. Also, the square and honeycomb lattices, with multiple atomic orbitals on each lattice site, can host TFBs [3, 11, 30-32]. On the other hand, however, very few electronic FB materials [33-42] have been identified, either experimentally or computationally, and realizations of the known FB lattice models are rather limited in general, with only a few examples such as the breathing-Kagome lattice in $Fe_3Sn_2$ [41], and the coloring-triangle lattice in Cu-dicyanobenzene monolayer [42] presented so far. This, apparently, has severely hindered the experimental realization of eluding FB properties. It is worth noting that the experimental discovery of FB-associated superconductivity in twisted bilayer graphene [43] has generated a lot of excitement recently, and a surprising ferromagnetic covalent-organic framework without transition-metal atoms [44] has been shown to be originated from a FB [10]. Therefore, computational design and identification of new FB materials, already exist or to be fabricated, is highly desirable to significantly advance the study of TFB physics, materials, and devices.

On the other hand, the recent establishment of materials database has enabled a high-throughput screening approach to discovering new materials in batches. For example, all the 3D nonmagnetic topological crystals are screened for compiling a complete catalogue of topological materials [45-47]; various nontrivial magnetic crystals are identified from 2D and 3D materials databases [48, 49]. Here, by screening the 2DMatPedia database [50] for 2D crystalline materials, we have identified fifteen monolayer atomic crystals hosting TFBs near the Fermi level, which cover six different FB lattice models. Among them, the Kagome sublattice of O in $TiO_2$ supports a spin-unpolarized TFB, and that of V in ferromagnetic $V_3F_8$ supports a spin-polarized TFB. Monolayer $Nb_3TeCl_7$ and its counterparts from element substitution, whose layered 3D structures were already synthesized over two decades ago, are found to be breathing-Kagome crystals with possible high-order topology. The family of monolayer $III_2VI_3$ compounds exhibit a TFB representing the coloring-triangle lattice model. $ReF_3$, $MnF_3$ and $MnBr_3$ are all predicted to be diatomic-Kagome crystals, with TFB transitions induced by atomic substitution. $HgF_2$, $CdF_2$ and $ZnF_2$ are discovered to host dual TFBs in the diamond-octagon lattice. Overall, some existing simple line-graph and generalized line-graph FB lattice models have been identified with several candidate materials.

*Screening procedure*. Our screening for 2D materials having TFBs starts with the database of 2DMatPedia, which contains electronic band structure without SOC for ~ 5300 monolayer atomic crystals [50], as shown in Fig. 1. The bandwidth $w$ of eleven bands (chosen for FB searching in this work) around the Fermi level is calculated, from which 354 nonmagnetic and 242 magnetic materials are revealed to possess bands whose $w$ is less than 50 meV. We note that the choice of this bandwidth threshold could be somewhat arbitrary, and we used a value that lies in the typical range of SOC-induced energy gap in realistic solid materials. In general, using a larger (smaller) value one would find more (less) FB materials from screening but a higher (lower) percentage of trivial ones. Next, if the narrow band is identified with a degenerate point with another dispersive band, its host material is selected, amounting to 27 candidate TFB materials [see Table S1 and S2 in Supplemental Material (SM) [51]]. Finally, these 27 candidates were further double checked by *ab-initio* calculations with high precision and lattice model analysis (see computational methods in SM), which confirmed 15 TFB materials (9 nonmagnetic and 6 magnetic) representing respectively six different TFB lattice models, as listed in Table 1. In the following, we choose the representative examples to show ideal TFBs discovered in (i) Kagome lattice, (ii) related breathing-Kagome, coloring-triangle, and diatomic-Kagome lattices, (iii) diamond-octagon lattice, and (iv) $d$-orbital graphene lattice.

*Kagome crystals with the ideal TFB*. The well-known Kagome lattice consists of three sites in a unit cell, exhibiting two Dirac bands touched with a TFB [Fig. 2(a), see tight-binding Hamiltonian in SM]. As shown in Fig. 3(a), the monolayer $TiO_2$, a widely-studied compound, has O atoms forming an upper and lower Kagome layer bridged by O atoms in the middle. It turns out to host a typical Kagome band structure from $p$ orbitals of O atoms, with a well separated FB right below the Fermi level while the Dirac bands mix with other trivial bands. As expected, an isolated TFB arises upon a nontrivial gap opening in the presence of SOC [Fig. S1(a) in SM]. There are also two additional sets of Kagome bands far below the Fermi level arising from O 2$s$ orbitals in the upper and lower Kagome layer, respectively (Fig. S2), sharing a doubly degenerate TFB. Another nonmagnetic Kagome crystal is found in monolayer $BaYSn_4O_7$, featuring a TFB around the Fermi level arising from Sn atoms sitting on a Kagome lattice [Fig. S3(a)]. Previously, Kagome bands have been mostly shown in metal-organic and covalent-organic frameworks [35-39]. Here we discover the unknown inorganic 2D materials hosting well-separated Kagome bands.

Spin-polarized TFBs are discovered in magnetic Kagome crystals with transition-metal elements. As shown in Fig. 3(b), three V atoms in monolayer $V_3F_8$ (shown to be stable in literature [55]), with magnetic moment $M = 7$ $\mu_B$, sit in a Kagome sublattice. Naturally, spin-up Kagome bands from $d$ orbitals of vanadium atoms arise, with the TFB lying close to the Fermi level based on electron counting. With SOC, a topological gap opens to isolate a Chern TFB [Fig. S1(b)], affording an intriguing possibility of exploring fractional quantum anomalous Hall effect. Another ferromagnetic Kagome material is found in monolayer $Li_2Fe_3F_8$, where the Kagome sublattice consists of Fe atoms each with $M = 4$ $\mu_B$ [Fig. S3(b)]. The above-mentioned Kagome materials are all intrinsic, without the need of doping, and ideal, without band overlapping with other trivial bands, superior over previous computationally- and experimentally-identified electronic Kagome metals [40, 56-60]. For the Kagome materials of $TiO_2$ and $V_3F_8$, their SOC-induced gap opening between the TFB and dispersive bands (Fig. S1) confirms the gapped TFB to possess a nonzero (spin) Chern invariant [4]. Similarly, the TFBs in other identified 2D materials are checked, where the gap is larger for materials with heavier atoms having stronger SOC.

*Breathing-Kagome crystals with potential second-order topology*. When one triangle in Kagome lattice shrinks and the other one expands, a so-called breathing-Kagome lattice model is constructed as a generalized line graph of hexagonal lattice [Fig. 2(b)]. This breathing mode does not affect the TFB, but opens a gap at Dirac point to induce the second-order topological corner states [23-25, 61, 62]. As shown in Fig. 4, monolayer $Nb_3TeCl_7$ is discovered to have Nb atoms locating at the breathing-Kagome lattice sites, where Nb $d$ orbitals constitute a clean set of Kagome bands with a Dirac gap at the Fermi level. Such breathing-Kagome state is also discovered in 2D $Ta_3SBr_7$ with Ta atoms locating at the breathing-Kagome sites [Fig. S4(a)]. Furthermore, spin-polarized breathing-Kagome states with $M = 1$ $\mu_B$ are discovered in monolayer $Nb_3Cl_8$ with three Nb atoms sitting at three breathing-Kagome sites, respectively [Fig. S4(b)]. While magnetic interaction separates the spin-up and -down bands in energy space, all the features of breathing-Kagome bands remain intact. Compared with $Nb_3TeCl_7$, the ferromagnetism originates from substituting a Te atom with a Cl atom; compared with $V_3F_8$, the breathing deformation in $Nb_3Cl_8$ originates from substituting V atoms by Nb atoms.

*Coloring-triangle crystals with Kagome bands*. When one triangle in Kagome lattice rotates 30° clockwise and the other triangle 30° counterclockwise, a so-called coloring-triangle lattice, which is another generalized line graph of hexagonal lattice by an off-edge-center construction [Fig. 2(c)], forms to also host identical Kagome bands [28]. The lattice can also be viewed as a triangle lattice with part of nearest-neighbor (NN) hopping being blocked, so that its realization in real crystalline materials is supposed to be quite difficult. Surprisingly, we have overcome this difficulty "accidentally". As shown in Fig. 5(a), within each unit cell of monolayer $B_2S_3$, S atoms constitute a triangle sublattice while B atoms are located at the center of two S triangles. Consequently, via the bridging of B atoms, the NN hopping between those S atoms in the two triangles with B are much stronger than those S atoms without bridging B atoms. So, effectively this provides a unique mechanism to selectively block part of NN hopping in a triangle S lattice, as required by the construction of a coloring-triangle lattice. This is clearly confirmed by the prefect Kagome bands arising from S $p_z$ orbitals in $B_2S_3$ [Fig. 5(a)]. The Kagome bands are actually embedded in other bands from S $p_{x,y}$ orbitals in the equilibrium $B_2S_3$ (Fig. S5), but can be separated out by applying a small biaxial tensile strain [Fig. 5(b)], which simultaneously moves the TFB upward closer to the Fermi level. Deep-energy Kagome bands from S $3s$ orbitals manifest also the coloring-triangle lattice in monolayer $B_2S_3$ (Fig. S5).

Similar to $B_2S_3$, our search reveals that most of the monolayer $III_2VI_3$ compounds, made from elements in the III and VI main group, exhibit the ideal Kagome FBs satisfying the coloring-triangle model upon strain-induced band separation [Table S3, Fig. S6(a)]. An exception is monolayer $Tl_2Te_3$, for which the Kagome states disappear due to the destruction of the desired coloring-triangle hopping, but on the other hand, its strong SOC induces a large topological gap of 0.32 eV [Fig. S6(b)], affording a candidate for high-temperature topological insulator. Also, the FB is absent in monolayer $In_2Se_3$, $Tl_2S_3$, and $Tl_2Se_3$ with large atoms, because of the structure-induced destruction of the desired coloring-triangle hopping [see details in Fig. S7].

*TFBs evolution in diatomic-Kagome crystals*. When every Kagome site is replaced by a pair of lattice sites, a diatomic-Kagome lattice is formed, which is yet another generalized line graph of hexagonal graph constructed with copies of breathing-Kagome lattices [Fig. 2(d)], leading to intriguing evolution of TFBs and phase transitions [16, 27]. For example, with a small next-NN (NNN) hopping integral, two sets of Kagome bands coexist [blue lines in Fig. 2(d)]; as the NNN hopping becomes stronger, the bands evolve into a combination of Dirac bands and $p_{x,y}$-orbital hexagonal lattice bands [gray lines in Fig. 2(d)], labeled as (D; $p_{x,y}$) phase [11, 30]. Interestingly, our screening process leads to the discovery of three diatomic-Kagome crystals, exhibiting the TFB transitions as proposed in tight-binding models. As shown in Fig. 6(a), monolayer $ReF_3$ has Re atoms sitting on the diatomic-Kagome sites, resulting in the (D; $p_{x,y}$) bands mainly contributed from Re-$d$ orbitals. In contrast, the monolayer $MnF_3$ and $MnBr_3$ are ferromagnetic with $M = 4$ $\mu_B$ on each Mn atom, possessing two sets of spin-up Kagome and (D; $p_{x,y}$) bands, respectively [Fig. 6(b), 6(c)]. From $ReF_3$ to $MnF_3$, the (D; $p_{x,y}$) bands transform into two sets of Kagome bands, which in turn transform back into the (D; $p_{x,y}$) bands from $MnF_3$ to MnBr$_3$. This indicates that atomic substitution is an effective way to tune the hopping integrals in diatomic-Kagome lattices, providing a promising strategy for topological and FB engineering [63].

*Dual TFBs in diamond-octagon crystals*. Lieb lattice contains a sublattice of checkerboard lattice which is the line-graph of square lattice, and hence possesses FB. The line graph of Lieb lattice (dubbed as diamond-octagon lattice) also possesses FB [2, 29]. With four lattice sites in a square unit cell [Fig. 2(e)], when the NN hopping integral is equal to the NNN hopping integral, two perfect TFBs can appear, which are degenerate with one parabolic band in between, with band touching at Γ and M points, respectively. Due to the usual exponential decay of lattice hopping with distance in real materials, this peculiar model with equal hopping for both closer and farther sites appeared difficult to be realized. Interestingly, in monolayer HgF$_2$ (Fig. 7), two Hg atoms on the horizontal boundaries of the unit cell are located at different heights from the other two on vertical boundaries. This leads to an almost equal distance between the NN Hg-Hg sites in the same plane to the NNN Hg-Hg sites in the different planes, so as to satisfy the desired electron hopping condition prescribed in the above model [Fig. 2(e)], similar to the case of bilayer Ni(CH) for realizing a diatomic-Kagome lattice [16]. Consequently, monolayer HgF$_2$ exhibits two TFBs inside four bands which are mainly contributed by $s$ orbitals of mercury atoms. Also, this type of dual TFBs is found in monolayer CdF$_2$ and ZnF$_2$ (Fig. S8).

*Orbital-enabled TFBs in honeycomb crystals*. Beyond TFBs from line-graph construction, atomic/molecular orbitals in non-line-graph lattices can also be exploited to produce TFBs [3, 11, 30-32]. Organometallic framework [11, 30, 31] and bismuthene on semiconductor substrates [64, 65] have been predicted to realize the honeycomb lattice with TFBs from $p$ orbitals. Here, as shown in Fig. 8, we discover monolayer VF$_3$ has V-$d$-orbital TFBs and Dirac bands, which conforms to the recently proposed TFBs from $d_{xy}$ and $d_{x^2-y^2}$ orbitals [3], and Dirac bands from $d_{z^2}$ in a honeycomb lattice, respectively. The TFB is just below the Fermi level and spin-polarized, with $M = 2$ $\mu_B$ on each V atom. In contrast, a previous $sd^2$-orbital honeycomb lattice with TFB has been shown in W overlayer on a halogenated Si(111) surface invoking an electronic Kagome lattice from the coupled $s$ and $d$ orbitals [32].

In all the above cases, the screening condition for band width $w < 50$ meV is used (Fig. 1). One might relax this condition to discover more FB materials. For example, we also found the monolayer $II_1VII_2$ compounds, made from elements in the II and VII main groups, have the checkerboard and diamond-octagon FBs with larger bandwidth, where the former arises from an intriguing $p$-orbital configuration that is equivalent to the diagonal $p$-orbital orientation in a square lattice [3], such as $MgCl_2$ shown in Fig. S9 of SM. Totally, two monolayer crystals of $VF_3$ and $MgCl_2$ are identified to be non-line-graph materials having the FB near the Fermi level, while others are line-graph FB materials. Beyond the analysis by using tight-binding lattice models, the Wannier construction for FBs of the non-line-graph $VF_3$ and line-graph $V_3F_8$ is carried out, which again confirms our finding that they respectively conform to the Kagome and honeycomb FB model [see details in Fig. S10].

*Potential experimental fabrication and measurement.* We note that the 3D layered $Nb_3TeCl_7$ and $Ta_3SBr_7$ have been experimentally synthesized over twenty years ago [66, 67], making experimental observation of these material-specific breathing-Kagome states highly promising. The 2D film of breathing-Kagome crystal $Nb_3Cl_8$ with controllable thickness has been already realized in experiment [68], which calls for immediate experimental confirmation of its predicted TFB. Also, some of coloring-triangle crystals, such as $Ga_2S_3$, are potentially experimentally achievable [69-71]. For other 2D FB crystals predicted here, we expect that they can be exfoliated from layered bulk materials by the mechanical, electrical, or chemical method [72, 73], or can be grown on suitable prescreened substrates by the molecular beam epitaxy, or chemical vapor deposition technology [74, 75].

The TFBs predicted in this work, all around Fermi level, can be directly probed by angle resolved photoemission spectroscopy [76], where the momentum-space band dispersion can be measured. Also, the FBs can be probed by scanning tunneling spectroscopy (STS), where a sharp peak of density of states could be shown as a signature for FBs [40]; and the edge/corner states from electron topology can be directly imaged by the space-resolved STS. The measurement of quantum Hall conductance can demonstrate topological transport properties [77], which would potentially confirm the intriguing fractional quantum anomalous Hall effect from partially occupied TFBs [6-8].

*Conclusions*. By employing the data screening calculations on materials from a first-principles materials database, we have discovered fifteen inorganic 2D crystals (Table 1) with ideal TFBs representing six FB lattice models, which opens a new door towards experimental observation of TFBs in real materials, and exploration/substantiation of their intriguing physics and applications. In particular, some of these realistic 2D crystals have already been made in experiments, which will hopefully draw immediate attention. Moreover, the screening approach developed here for searching 2D FB crystals can be extended to systematically uncovering 3D FB materials from the well-established databases [78-81].

F. L. is supported by U.S. DOE-BES (Grant No. DE-FG02-04ER46148). S. M. and H. L. thank financial support from the National Natural Science Foundation of China (Grants No. 12025407 and 11934003), and Chinese Academy of Sciences (Grant No. XDB330301).

---

[*]smeng@iphy.ac.cn

[†]fliu@eng.utah.edu

[1] Z. Liu, F. Liu, and Y.-S. Wu, Exotic electronic states in the world of flat bands: From theory to material. *Chin. Phys. B* **23**, 077308 (2014).

[2] D.-S. Ma, Y. Xu, C. S. Chiu, N. Regnault, A. A. Houck, Z. Song, and B. A. Bernevig, Spin-Orbit-Induced Topological Flat Bands in Line and Split Graphs of Bipartite Lattices. *Phys. Rev. Lett.* **125**, 266403 (2020).

[3] H. Liu, G. Sethi, S. Meng, and F. Liu, Orbital Design of Flat Bands in Non-Line-Graph Lattices via Line-Graph Wavefunctions. *To be published.*

[4] J.-W. Rhim, and B.-J. Yang, Classification of flat bands according to the band-crossing singularity of Bloch wave functions. *Phys. Rev. B* **99**, 045107 (2019).

[5] D. L. Bergman, C. Wu, and L. Balents, Band touching from real-space topology in frustrated hopping models. *Phys. Rev. B* **78**, 125104 (2008).

[6] E. Tang, J. W. Mei, and X. G. Wen, High-temperature fractional quantum Hall states. *Phys. Rev. Lett.* **106**, 236802 (2011).

[7] T. Neupert, L. Santos, C. Chamon, and C. Mudry, Fractional quantum Hall states at zero magnetic field. *Phys. Rev. Lett.* **106**, 236804 (2011).

[8] K. Sun, Z. Gu, H. Katsura, and S. Das Sarma, Nearly flatbands with nontrivial topology. *Phys. Rev. Lett.* **106**, 236803 (2011).

[9] E. C. Stoner, Collective electron ferronmagnetism. *Proc. R. Soc. Lond. A* **165**, 372 (1938).

[10] W. Jiang, H. Huang, and F. Liu, A Lieb-like lattice in a covalent-organic framework and its Stoner ferromagnetism. *Nat. Commun.* **10**, 2207 (2019).

[11] C. Wu, D. Bergman, L. Balents, and S. Das Sarma, Flat bands and Wigner crystallization in the honeycomb optical lattice. *Phys. Rev. Lett.* **99**, 070401 (2007).

[12] B. Jaworowski, A. D. Güçlü, P. Kaczmarkiewicz, M. Kupczyński, P. Potasz, and A. Wójs, Wigner crystallization in topological flat bands. *New Journal of Physics* **20**, 063023 (2018).

[13] S. Miyahara, S. Kusuta, and N. Furukawa, BCS theory on a flat band lattice. *Physica C: Superconductivity* **460-462**, 1145 (2007).

[14] D. Jérome, T. M. Rice, and W. Kohn, Excitonic Insulator. *Physical Review* **158**, 462 (1967).

[15] G. Sethi, Y. Zhou, L. Zhu, L. Yang, and F. Liu, Flat-Bands-Enabled Triplet Excitonic Insulator in a Diatomic (Yin-Yang) Kagome Lattice. *arXiv:2102.08593*.

[16] Y. Zhou, G. Sethi, H. Liu, Z. Wang, and F. Liu, Excited quantum Hall effect: enantiomorphic flat bands in a Yin-Yang Kagome lattice. *arXiv:1908.03689*.

[17] A. Mielke, Ferromagnetic ground states for the Hubbard model on line graphs. *J. Phys. A - Math. Gen.* **24**, L73 (1991).

[18] A. Mielke, Ferromagnetism in the Hubbard model on line graphs and further considerations. *J. Phys. A - Math. Gen.* **24**, 3311 (1991).

[19] A. Mielke, Exact ground states for the Hubbard model on the Kagome lattice. *J. Phys. A - Math. Gen.* **25**, 4335 (1992).

[20] I. Syôzi, Statistics of Kagomé Lattice. *Progress of Theoretical Physics* **6**, 306 (1951).

[21] E. H. Lieb, Two theorems on the Hubbard model. *Phys. Rev. Lett.* **62**, 1201 (1989).

[22] H. Tasaki, From Nagaoka's Ferromagnetism to Flat-Band Ferromagnetism and Beyond. *Progress of Theoretical Physics* **99**, 489 (1998).

[23] M. Ezawa, Higher-Order Topological Insulators and Semimetals on the Breathing Kagome and Pyrochlore Lattices. *Phys. Rev. Lett.* **120**, 026801 (2018).

[24] H. Xue, Y. Yang, F. Gao, Y. Chong, and B. Zhang, Acoustic higher-order topological insulator on a kagome lattice. *Nat. Mater.* **18**, 108 (2019).

[25] A. Bolens, and N. Nagaosa, Topological states on the breathing kagome lattice. *Phys. Rev. B* **99**, 165414 (2019).

[26] K. Essafi, L. D. C. Jaubert, and M. Udagawa, Flat bands and Dirac cones in breathing lattices. *J. Phys. Condens. Matter* **29**, 315802 (2017).

[27] Y. Zhou, G. Sethi, C. Zhang, X. Ni, and F. Liu, Giant intrinsic circular dichroism of enantiomorphic flat Chern bands and flatband devices. *Phys. Rev. B* **102**, 125115 (2020).

[28] S. Zhang, M. Kang, H. Huang, W. Jiang, X. Ni, L. Kang, S. Zhang, H. Xu, Z. Liu, and F. Liu, Kagome bands disguised in a coloring-triangle lattice. *Phys. Rev. B* **99**, 100404(R) (2019).

[29] B. Pal, Nontrivial topological flat bands in a diamond-octagon lattice geometry. *Phys. Rev. B* **98**, 245116 (2018).

[30] C. Wu, and S. Das Sarma, px,y-orbital counterpart of graphene: Cold atoms in the honeycomb optical lattice. *Phys. Rev. B* **77**, 235107 (2008).

[31] Z. Liu, Z. F. Wang, J. W. Mei, Y. S. Wu, and F. Liu, Flat Chern band in a two-dimensional organometallic framework. *Phys. Rev. Lett.* **110**, 106804 (2013).

[32] M. Zhou, Z. Liu, W. Ming, Z. Wang, and F. Liu, sd(2) Graphene: Kagome band in a hexagonal lattice. *Phys. Rev. Lett.* **113**, 236802 (2014).

[33] X. Zhang, and M. Zhao, Robust half-metallicity and topological aspects in two-dimensional Cu-TPyB. *Sci. Rep.* **5**, 14098 (2015).

[34] O. J. Silveira, S. S. Alexandre, and H. Chacham, Electron States of 2D Metal–Organic and Covalent–Organic Honeycomb Frameworks: Ab Initio Results and a General Fitting Hamiltonian. *J. Phys. Chem. C* **120**, 19796 (2016).

[35] Z. F. Wang, N. Su, and F. Liu, Prediction of a two-dimensional organic topological insulator. *Nano Lett.* **13**, 2842 (2013).

[36] M. G. Yamada, T. Soejima, N. Tsuji, D. Hirai, M. Dincă, and H. Aoki, First-principles design of a half-filled flat band of the kagome lattice in two-dimensional metal-organic frameworks. *Phys. Rev. B* **94**, 081102(R) (2016).

[37] L. Z. Zhang, Z. F. Wang, B. Huang, B. Cui, Z. Wang, S. X. Du, H. J. Gao, and F. Liu, Intrinsic Two-Dimensional Organic Topological Insulators in Metal-Dicyanoanthracene Lattices. *Nano Lett.* **16**, 2072 (2016).

[38] X. Zhang, Z. Wang, M. Zhao, and F. Liu, Tunable topological states in electron-doped HTT-Pt. *Phys. Rev. B* **93**, 165401 (2016).

[39] O. J. Silveira, and H. Chacham, Electronic and spin-orbit properties of the kagome MOF family $M_3(1,2,5,6,9,10$-triphenylenehexathiol$)_2$ (M = Ni, Pt, Cu and Au). *J. Phys. Condens. Matter* **29**, 09LT01 (2017).

[40] Z. Li, J. Zhuang, L. Wang, H. Feng, Q. Gao, X. Xu, W. Hao, X. Wang, C. Zhang, K. Wu, S. X. Dou, L. Chen, Z. Hu, and Y. Du, Realization of flat band with possible nontrivial topology in electronic Kagome lattice. *Sci. Adv.* **4**, eaau4511 (2018).

[41] H. Tanaka, Y. Fujisawa, K. Kuroda, R. Noguchi, S. Sakuragi, C. Bareille, B. Smith, C. Cacho, S. W. Jung, T. Muro, Y. Okada, and T. Kondo, Three-dimensional electronic structure in ferromagnetic $Fe_3Sn_2$ with breathing kagome bilayers. *Phys. Rev. B* **101**, 161114(R) (2020).

[42] Y. Gao, Y.-Y. Zhang, J.-T. Sun, L. Zhang, S. Zhang, and S. Du, Quantum anomalous Hall effect in two-dimensional Cu-dicyanobenzene coloring-triangle lattice. *Nano Research* **13**, 1571 (2020).

[43] Y. Cao, V. Fatemi, S. Fang, K. Watanabe, T. Taniguchi, E. Kaxiras, and P. Jarillo-Herrero, Unconventional superconductivity in magic-angle graphene superlattices. *Nature* **556**, 43 (2018).

[44] E. Jin, M. Asada, Q. Xu, S. Dalapati, M. A. Addicoat, M. A. Brady, H. Xu, T. Nakamura, T. Heine, Q. Chen, and D. Jiang, Two-dimensional $sp^2$ carbon–conjugated covalent organic frameworks. *Science* **357**, 673 (2017).

[45] F. Tang, H. C. Po, A. Vishwanath, and X. Wan, Comprehensive search for topological materials using symmetry indicators. *Nature* **566**, 486 (2019).

[46] M. G. Vergniory, L. Elcoro, C. Felser, N. Regnault, B. A. Bernevig, and Z. Wang, A complete catalogue of high-quality topological materials. *Nature* **566**, 480 (2019).

[47] T. Zhang, Y. Jiang, Z. Song, H. Huang, Y. He, Z. Fang, H. Weng, and C. Fang, Catalogue of topological electronic materials. *Nature* **566**, 475 (2019).

[48] H. Liu, J. T. Sun, M. Liu, and S. Meng, Screening Magnetic Two-Dimensional Atomic Crystals with Nontrivial Electronic Topology. *J. Phys. Chem. Lett.* **9**, 6709 (2018).

[49] Y. Xu, L. Elcoro, Z.-D. Song, B. J. Wieder, M. G. Vergniory, N. Regnault, Y. Chen, C. Felser, and B. A. Bernevig, High-throughput calculations of magnetic topological materials. *Nature* **586**, 702 (2020).

[50] J. Zhou, L. Shen, M. D. Costa, K. A. Persson, S. P. Ong, P. Huck, Y. Lu, X. Ma, Y. Chen, H. Tang, and Y. P. Feng, 2DMatPedia, an open computational database of two-dimensional materials from top-down and bottom-up approaches. *Sci. Data* **6**, 86 (2019).

[51] See Supplemental Material at http://link.aps.org/supplemental/xxx for details about computational methods, list of TFB materials, SOC bands, tight-binding modeling, and bands of TFB materials not shown in main text, which includes Refs. [3, 52-54].

[52] G. Kresse, and J. Furthmuller, Efficient iterative schemes for ab initio total-energy calculations using a plane-wave basis set. *Phys. Rev. B* **54**, 11169 (1996).

[53] J. P. Perdew, K. Burke, and M. Ernzerhof, Generalized Gradient Approximation Made Simple. *Phys. Rev. Lett.* **77**, 3865 (1996).

[54] A. A. Mostofi, J. R. Yates, G. Pizzi, Y.-S. Lee, I. Souza, D. Vanderbilt, and N. Marzari, An updated version of wannier90: A tool for obtaining maximally-localised Wannier functions. *Comput. Phys. Commun.* **185**, 2309 (2014).

[55] H. Xiao, X. Wang, R. Wang, L. Xu, S. Liang, and C. Yang, Intrinsic magnetism and biaxial strain tuning in two-dimensional metal halides V3X8 (X = F, Cl, Br, I) from first principles and Monte Carlo simulation. *Phys Chem Chem Phys* **21**, 11731 (2019).

[56] M. Kang, S. Fang, L. Ye, H. C. Po, J. Denlinger, C. Jozwiak, A. Bostwick, E. Rotenberg, E. Kaxiras, J. G. Checkelsky, and R. Comin, Topological flat bands in frustrated kagome lattice CoSn. *Nat. Commun.* **11**, 4004 (2020).

[57] B. R. Ortiz, S. M. L. Teicher, Y. Hu, J. L. Zuo, P. M. Sarte, E. C. Schueller, A. M. M. Abeykoon, M. J. Krogstad, S. Rosenkranz, R. Osborn, R. Seshadri, L. Balents, J. He, and S. D. Wilson, CsV3Sb5: A Z2 Topological Kagome Metal with a Superconducting Ground State. *Phys. Rev. Lett.* **125**, 247002 (2020).

[58] L. Ye, M. Kang, J. Liu, F. von Cube, C. R. Wicker, T. Suzuki, C. Jozwiak, A. Bostwick, E. Rotenberg, D. C. Bell, L. Fu, R. Comin, and J. G. Checkelsky, Massive Dirac fermions in a ferromagnetic kagome metal. *Nature* **555**, 638 (2018).

[59] Z. Lin, J. H. Choi, Q. Zhang, W. Qin, S. Yi, P. Wang, L. Li, Y. Wang, H. Zhang, Z. Sun, L. Wei, S. Zhang, T. Guo, Q. Lu, J. H. Cho, C. Zeng, and Z. Zhang, Flatbands and Emergent Ferromagnetic Ordering in Fe3Sn2 Kagome Lattices. *Phys. Rev. Lett.* **121**, 096401 (2018).

[60] M. Kang, L. Ye, S. Fang, J. S. You, A. Levitan, M. Han, J. I. Facio, C. Jozwiak, A. Bostwick, E. Rotenberg, M. K. Chan, R. D. McDonald, D. Graf, K. Kaznatcheev, E. Vescovo, D. C. Bell, E. Kaxiras, J. van den Brink, M. Richter, M. Prasad Ghimire, J. G. Checkelsky, and R. Comin, Dirac fermions and flat bands in the ideal kagome metal FeSn. *Nat. Mater.* **19**, 163 (2020).

[61] X. Ni, M. Weiner, A. Alu, and A. B. Khanikaev, Observation of higher-order topological acoustic states protected by generalized chiral symmetry. *Nat. Mater.* **18**, 113 (2019).

[62] S. N. Kempkes, M. R. Slot, J. J. van den Broeke, P. Capiod, W. A. Benalcazar, D. Vanmaekelbergh, D. Bercioux, I. Swart, and C. Morais Smith, Robust zero-energy modes in an electronic higher-order topological insulator. *Nat. Mater.* **18**, 1292 (2019).

[63] B. Huang, M. Yoon, B. G. Sumpter, S. H. Wei, and F. Liu, Alloy Engineering of Defect Properties in Semiconductors: Suppression of Deep Levels in Transition-Metal Dichalcogenides. *Phys. Rev. Lett.* **115**, 126806 (2015).

[64] M. Zhou, W. Ming, Z. Liu, Z. Wang, P. Li, and F. Liu, Epitaxial growth of large-gap quantum spin Hall insulator on semiconductor surface. *Proceedings of the National Academy of Sciences* **111**, 14378 (2014).

[65] M. Zhou, W. Ming, Z. Liu, Z. Wang, Y. Yao, and F. Liu, Formation of quantum spin Hall state on Si surface and energy gap scaling with strength of spin orbit coupling. *Sci. Rep.* **4**, 7102 (2014).

[66] G. J. Miller, Solid state chemistry of Nb3Cl8: Nb3TeCl7, mixed crystal formation, and intercalation. *Journal of Alloys and Compounds* **217**, 5 (1995).

[67] M. Smith, and G. J. Miller, Ta3SBr7 — A New Structure Type in the M3QX7 Family (M = Nb, Ta; Q = S, Se, Te; X = Cl, Br, I). *Journal of Solid State Chemistry* **140**, 226 (1998).

[68] J. Yoon, E. Lesne, K. Sklarek, J. Sheckelton, C. Pasco, S. S. P. Parkin, T. M. McQueen, and M. N. Ali, Anomalous thickness-dependent electrical conductivity in van der Waals layered transition metal halide, Nb3Cl8. *Journal of Physics: Condensed Matter* **32**, 304004 (2020).

[69] M. M. Y. A. Alsaif, N. Pillai, S. Kuriakose, S. Walia, A. Jannat, K. Xu, T. Alkathiri, M. Mohiuddin, T. Daeneke, K. Kalantar-Zadeh, J. Z. Ou, and A. Zavabeti, Atomically Thin Ga2S3 from Skin of Liquid Metals for Electrical, Optical, and Sensing Applications. *ACS Applied Nano Materials* **2**, 4665 (2019).

[70] X. Wang, Y. Sheng, R. J. Chang, J. K. Lee, Y. Zhou, S. Li, T. Chen, H. Huang, B. F. Porter, H. Bhaskaran, and J. H. Warner, Chemical Vapor Deposition Growth of Two-Dimensional Monolayer Gallium Sulfide Crystals Using Hydrogen Reduction of Ga2S3. *ACS Omega* **3**, 7897 (2018).

[71] L. Hu, and X. Huang, Peculiar electronic, strong in-plane and out-of-plane second harmonic generation and piezoelectric properties of atom-thick α-M2X3 (M = Ga, In; X = S, Se): role of spontaneous electric dipole orientations. *RSC Advances* **7**, 55034 (2017).

[72] K. S. Novoselov, A. Mishchenko, A. Carvalho, and A. H. Castro Neto, 2D materials and van der Waals heterostructures. *Science* **353**, aac9439 (2016).

[73] K. S. Novoselov, A. K. Geim, S. V. Morozov, D. Jiang, Y. Zhang, S. V. Dubonos, I. V. Grigorieva, and A. A. Firsov, Electric Field Effect in Atomically Thin Carbon Films. *Science* **306**, 666 (2004).

[74] J. Deng, B. Xia, X. Ma, H. Chen, H. Shan, X. Zhai, B. Li, A. Zhao, Y. Xu, W. Duan, S. C. Zhang, B. Wang, and J. G. Hou, Epitaxial growth of ultraflat stanene with topological band inversion. *Nat. Mater.* **17**, 1081 (2018).

[75] J. Zhou, J. Lin, X. Huang, Y. Zhou, Y. Chen, J. Xia, H. Wang, Y. Xie, H. Yu, J. Lei, D. Wu, F. Liu, Q. Fu, Q. Zeng, C. H. Hsu, C. Yang, L. Lu, T. Yu, Z. Shen, H. Lin, B. I. Yakobson, Q. Liu, K. Suenaga, G. Liu, and Z. Liu, A library of atomically thin metal chalcogenides. *Nature* **556**, 355 (2018).

[76] J. Kim, S. S. Baik, S. H. Ryu, Y. Sohn, S. Park, B. G. Park, J. Denlinger, Y. Yi, H. J. Choi, and K. S. Kim, Observation of tunable band gap and anisotropic Dirac semimetal state in black phosphorus. *Science* **349**, 723 (2015).

[77] C.-Z. Chang, J. Zhang, X. Feng, J. Shen, Z. Zhang, M. Guo, K. Li, Y. Ou, P. Wei, L.-L. Wang, Z.-Q. Ji, Y. Feng, S. Ji, X. Chen, J. Jia, X. Dai, Z. Fang, S.-C. Zhang, K. He, Y. Wang, L. Lu, X.-C. Ma, and Q.-K. Xue, Experimental Observation of the Quantum Anomalous Hall Effect in a Magnetic Topological Insulator. *Science* **340**, 167 (2013).

[78] A. Jain, S. P. Ong, G. Hautier, W. Chen, W. D. Richards, S. Dacek, S. Cholia, D. Gunter, D. Skinner, G. Ceder, and K. A. Persson, Commentary: The Materials Project: A materials genome approach to accelerating materials innovation. *APL Materials* **1**, 011002 (2013).

[79] S. Curtarolo, W. Setyawan, G. L. W. Hart, M. Jahnatek, R. V. Chepulskii, R. H. Taylor, S. Wang, J. Xue, K. Yang, O. Levy, M. J. Mehl, H. T. Stokes, D. O. Demchenko, and D. Morgan, AFLOW: An automatic framework for high-throughput materials discovery. *Comp. Mater. Sci.* **58**, 218 (2012).

[80] C. Draxl, and M. Scheffler, NOMAD: The FAIR Concept for Big-Data-Driven Materials Science. *arXiv:1805.05039*.

[81] G. Pizzi, A. Cepellotti, R. Sabatini, N. Marzari, and B. Kozinsky, AiiDA: automated interactive infrastructure and database for computational science. *Comp. Mater. Sci.* **111**, 218 (2016).

**Figures and Tables**

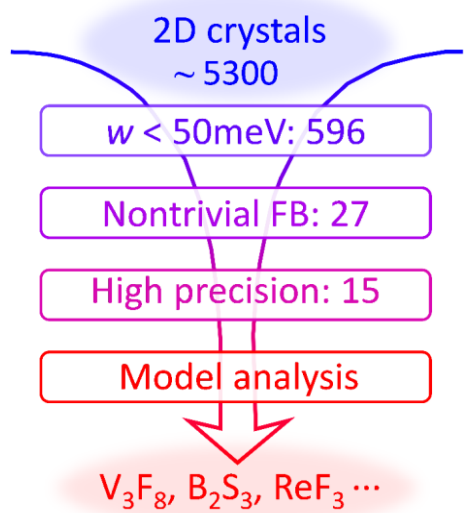

FIG. 1. Flowchart of computational screening for 2D crystals with ideal TFBs.

Table 1. The discovered monolayer atomic crystals with ideal TFBs.

| Model | Nonmagnetic | Magnetic | Space group |
|---|---|---|---|
| Kagome | $TiO_2$, $BaYSn_4O_7$ | $V_3F_8$, $Li_2Fe_3F_8$ | P-3m1 |
| Breathing-Kagome | $Nb_3TeCl_7$, $Ta_3SBr_7$ | $Nb_3Cl_8$ | |
| Coloring-triangle | $B_2S_3$ ($III_2VI_3$) | — | P-62m |
| Diatomic-Kagome | $ReF_3$ | $MnF_3$, $MnBr_3$ | P-3m1 |
| Honeycomb | — | $VF_3$ | |
| Diamond-octagon | $CdF_2$, $ZnF_2$, $HgF_2$ | — | P-4m2 |

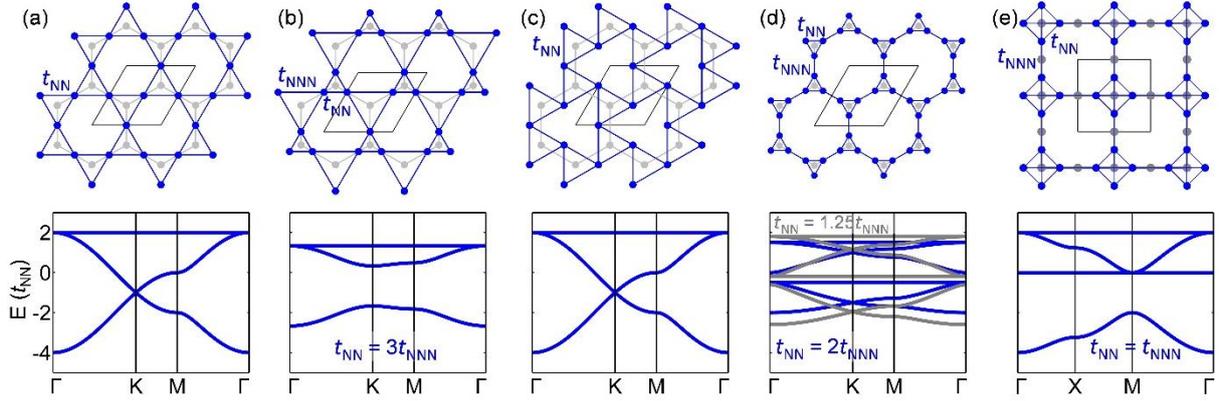

FIG. 2. Lattice structure and tight-binding bands of (a) Kagome lattice, and its derivatives including (b) breathing-Kagome, (c) coloring-triangle, (d) diatomic-Kagome lattices, and (e) diamond-octagon lattice. $t_{NN}$ and $t_{NNN}$ represent the NN and NNN hopping integral, respectively. Gray and blue dots/lines in upper panels show the original lattices and their (generalized) line-graph lattices, respectively. For standard construction [Fig. 2(a) and 2(e)], a line graph takes edge centers of the original graph as its vertices; while for generalized construction, a line graph takes off-edge-center positions, along the edge [Fig. 2(b)] or off the edge [Fig. 2(c)], as its vertices. Figure 2(d) can be viewed as two copies of the generalized line graph in Fig. 2(b), leading to dual FBs.

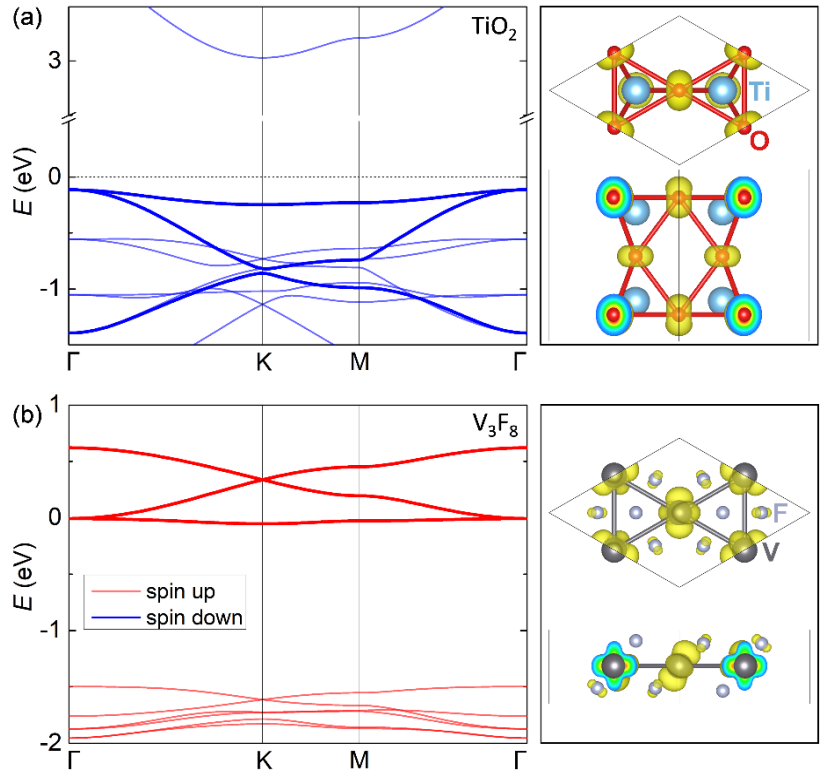

FIG. 3. Ideal Kagome bands in monolayer Kagome atomic crystals. (a) Oxygen atoms in monolayer $TiO_2$ form upper and lower Kagome layers connected by middle sites, with titanium atoms locating at the center of oxygen tetrahedrons. The electronic structure exhibits obvious Kagome bands (bold lines in left panel), and the charge of TFB near the Fermi level is contributed by $p$ orbitals of oxygen atoms (right panel). (b) Vanadium atoms in monolayer $V_3F_8$ form the perfect Kagome lattice, resulting in the typical Kagome bands (bold lines in left panel) with a TFB exactly locating at Fermi level. The $d$ orbitals of vanadium atoms contribute to the electron charge of the spin-up Kagome bands (right panel).

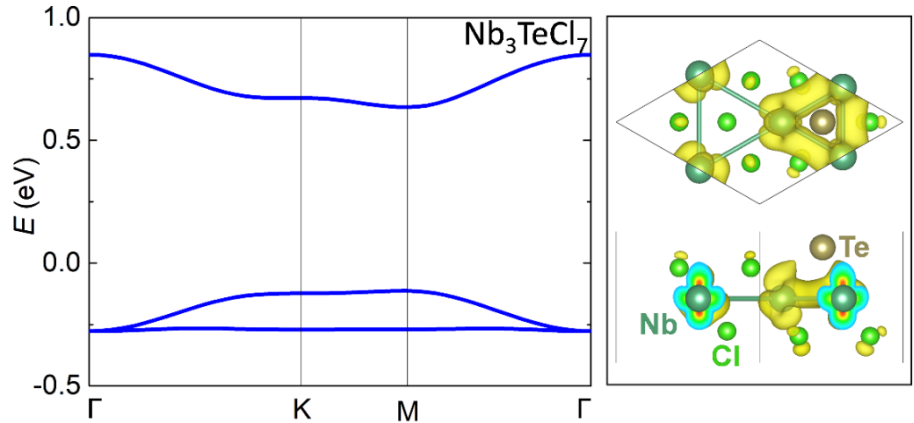

FIG. 4. Left: The electronic structure of monolayer $Nb_3TeCl_7$ with niobium atoms sitting on a breathing-Kagome lattice. Right: Electron charge distribution contributing to the breathing-Kagome bands with a gapped Dirac cone in the left.

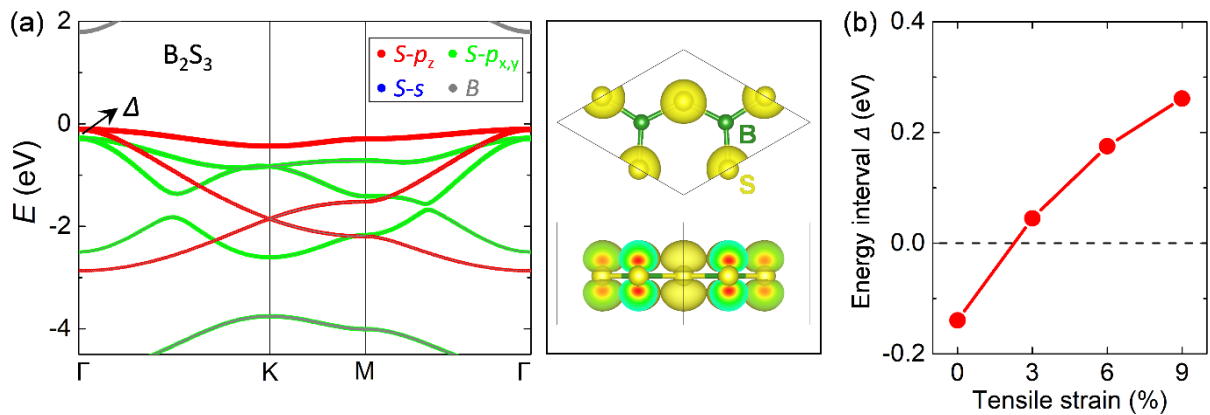

FIG. 5. (a) Electronic structure and orbital composition of monolayer $B_2S_3$ with +6% tensile strain (left panel). $\Delta$ marks the $\Gamma$-point energy interval between $p_z$ and $p_{x,y}$ bands just below the Fermi level. The $p_z$ orbitals of sulphur atoms, occupying the sites of a coloring-triangle lattice (right panel), contribute to the TFB of Kagome bands near the Fermi level. (b) The variation of $\Delta$ with external tensile strain.

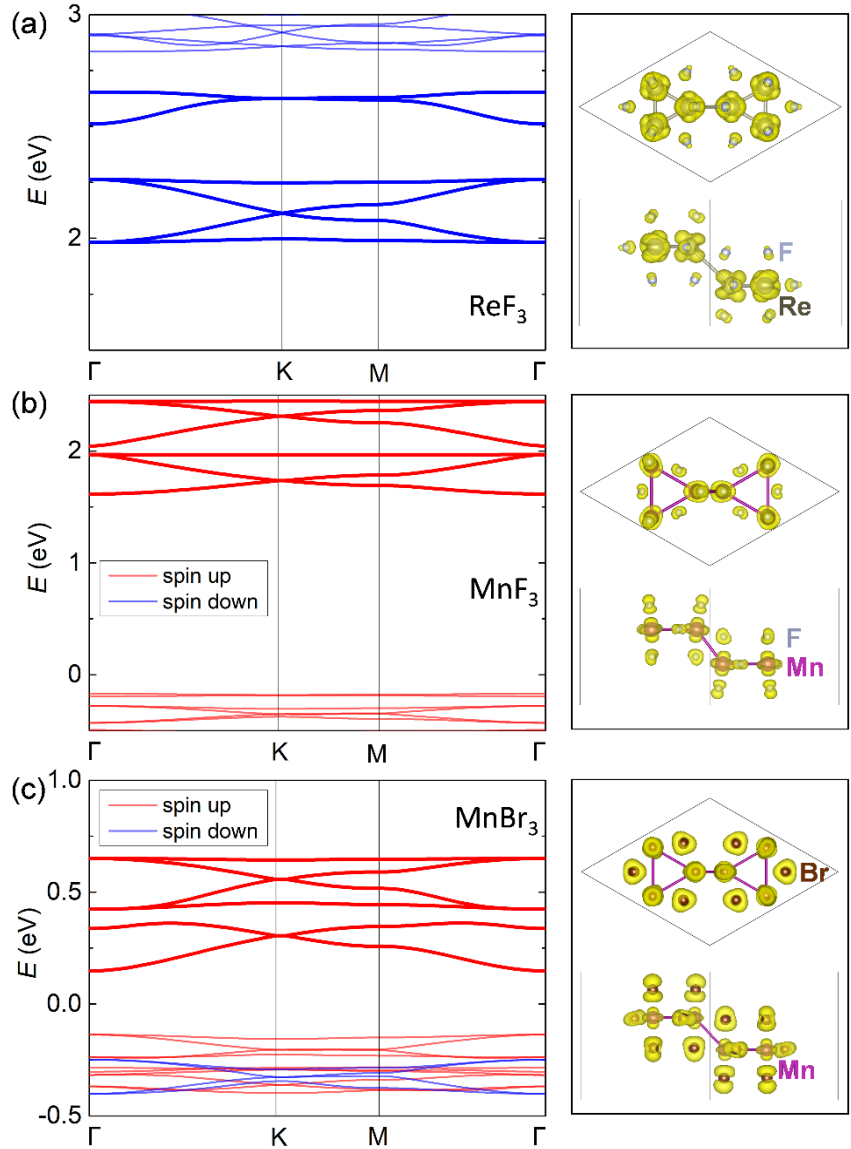

FIG. 6. Atomic and electronic structure for diatomic-Kagome crystals of (a) ReF$_3$, (b) MnF$_3$, and (c) MnBr$_3$. Left: Band structure with the diatomic-Kagome bands highlighted by thick lines. Right: The distribution of electron charge density for the diatomic-Kagome bands highlighted in the left.

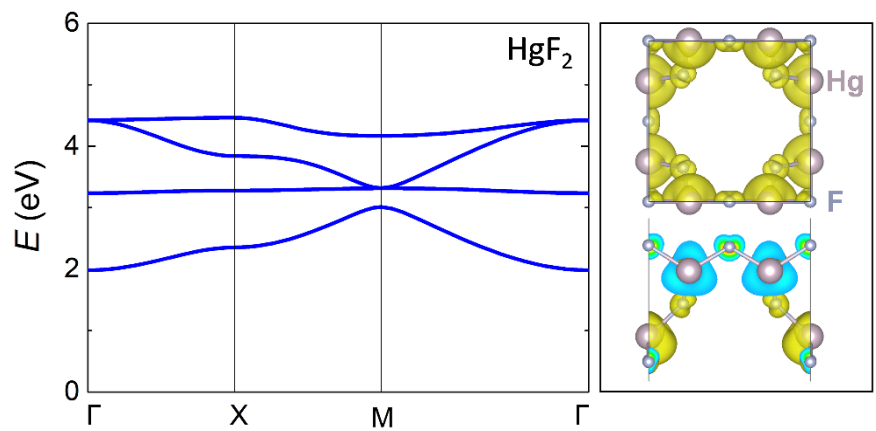

FIG. 7. Dual TFBs in monolayer HgF$_2$ with a diamond-octagon lattice. Left: Electron band structure. Right: Change density plot showing the bands in the left are mainly contributed from *s* orbitals of mercury atoms.

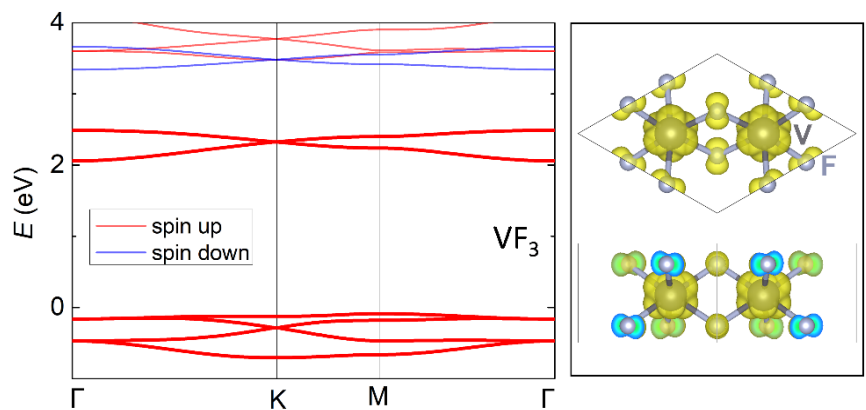

FIG. 8. Honeycomb TFB crystal of monolayer VF$_3$. Left: Band structure. Right: Electron charge density distribution contributing to the highlighted bands in the left.

Supplemental Material for

# Screening 2D materials with topological flat bands

Hang Liu[1,2,3], Sheng Meng[2,3,*], and Feng Liu[1,†]

[1] *Department of Materials Science and Engineering, University of Utah, Salt Lake City, Utah 84112, USA*

[2] *Songshan Lake Materials Laboratory, Dongguan, Guangdong 523808, People's Republic of China*

[3] *Beijing National Laboratory for Condensed Matter Physics and Institute of Physics, Chinese Academy of Sciences, Beijing 100190, People's Republic of China*

## Computational methods

I. First-principles calculations

The first-principles calculations are performed by using the Vienna Ab-initio Simulation Package [1]. We adopt the projector-augmented wave pseudopotential and Perdew-Burke-Ernzerhof functional [2] to describe the electrons-electrons and electrons-nucleus interactions, respectively. The energy cutoff of plane-wave basis set is set as 500 eV, and the first Brillouin zone is sampled by $9 \times 9 \times 1$ Gamma centered $k$-mesh. The convergence condition of electronic self-consistent loop is $10^{-6}$ eV. The correction from on-site Coulomb (U) and spin-spin (J) interactions is considered for spin-polarized materials. The Wannier fitting for some identified materials was done by Wannier90 package [3].

*smeng@iphy.ac.cn

†fliu@eng.utah.edu

[1] G. Kresse, and J. Furthmuller, Efficient iterative schemes for ab initio total-energy calculations using a plane-wave basis set. *Phys. Rev. B* **54**, 11169 (1996).

[2] J. P. Perdew, K. Burke, and M. Ernzerhof, Generalized Gradient Approximation Made Simple. *Phys. Rev. Lett.* **77**, 3865 (1996).

[3] A. A. Mostofi, J. R. Yates, G. Pizzi, Y.-S. Lee, I. Souza, D. Vanderbilt, and N. Marzari, An updated version of wannier90: A tool for obtaining maximally-localised Wannier functions. *Comput. Phys. Commun.* **185**, 2309 (2014).

[4] H. Liu, G. Sethi, S. Meng, and F. Liu, Orbital Design of Flat Bands in Non-Line-Graph Lattices via Line-Graph Wavefunctions. *To be published.*

## II. Tight-binding Hamiltonians

1. Kagome lattice.

$$H = 2t_{NN} \begin{pmatrix} 0 & \cos k_1 & \cos k_2 \\ & 0 & \cos k_3 \\ \dagger & & 0 \end{pmatrix} \quad (S1)$$

2. Breathing-Kagome lattice.

$$H = \begin{pmatrix} 0 & t_{NN}e^{ik_1/3} + t_{NNN}e^{-i2k_1/3} & t_{NN}e^{ik_2/3} + t_{NNN}e^{-i2k_2/3} \\ & 0 & t_{NN}e^{ik_3/3} + t_{NNN}e^{-i2k_3/3} \\ \dagger & & 0 \end{pmatrix} \quad (S2)$$

3. Coloring-triangle lattice.

$$H = t_{NN} \begin{pmatrix} 0 & e^{ik_1} + e^{ik_2} & e^{-ik_1} + e^{ik_3} \\ & 0 & e^{ik_2} + e^{-ik_3} \\ \dagger & & 0 \end{pmatrix} \quad (S3)$$

4. Diatomic-Kagome lattice.

$$H = \begin{pmatrix} 0 & t_{NN}e^{ik_{A_1A_2}} & t_{NN}e^{ik_{A_1A_3}} & t_{NNN}e^{ik_{A_1B_1}} & 0 & 0 \\ & 0 & t_{NN}e^{ik_{A_2A_3}} & 0 & t_{NNN}e^{ik_{A_2B_2}} & 0 \\ & & 0 & 0 & 0 & t_{NNN}e^{ik_{A_3B_3}} \\ & & & 0 & t_{NN}e^{ik_{B_1B_2}} & t_{NN}e^{ik_{B_1B_3}} \\ & \dagger & & & 0 & t_{NN}e^{ik_{B_2B_3}} \\ & & & & & 0 \end{pmatrix} \quad (S4)$$

$A_n$, $B_n$ (n = 1,2,3) labels six sites in diatomic-Kagome lattice; $k_{A_1A_2} = \boldsymbol{k} \cdot \overrightarrow{A_1A_2}$.

5. Diamond-octagon lattice.

$$H = \begin{pmatrix} 0 & 2t_{NNN}\cos\dfrac{k_1}{2} & t_{NN}e^{i\frac{k_2-k_1}{4}} & t_{NN}e^{i\frac{-k_2-k_1}{4}} \\ & 0 & t_{NN}e^{i\frac{k_2+k_1}{4}} & t_{NN}e^{i\frac{-k_2+k_1}{4}} \\ & & 0 & 2t_{NNN}\cos\dfrac{k_2}{2} \\ & & & 0 \end{pmatrix} \quad (S5)$$

Note: $k_n = \boldsymbol{k} \cdot \boldsymbol{a}_n$ with neighboring vector $\boldsymbol{a}_n$ of lattices in Fig. 2.

Table S1. 14 nonmagnetic crystals with possible TFBs.

| No. | Mat-ID | Formula | Bandwidth (meV) | Model |
|---|---|---|---|---|
| 1 | 2dm-3785 | $Nb_3TeCl_7$ | 11.8 | Breathing-Kagome |
| 2 | 2dm-138 | $CdF_2$ | 13.7 | Diamond-octagon |
| 3 | 2dm-5108 | $TiO_2$ | 14.6 | Kagome |
| 4 | 2dm-1537 | $ReF_3$ | 16.9 | Diatomic-Kagome |
| 5 | 2dm-4110 | $BaYSn_4O_7$ | 29.5 | Kagome |
| 6 | 2dm-5290 | $Y_3I_7O$ | 32.3 | − |
| 7 | 2dm-1984 | $ZnF_2$ | 41.7 | Diamond-octagon |
| 8 | 2dm-454 | $S_4O_9$ | 45.0 | − |
| 9 | 2dm-263 | $BiCl_3$ | 45.6 | − |
| 10 | 2dm-5348 | $Ta_3SBr_7$ | 45.8 | Breathing-Kagome |
| 11 | 2dm-1888 | $HgF_2$ | 46.2 | Diamond-octagon |
| 12 | 2dm-691 | $RhF_3$ | 46.8 | − |
| 13 | 2dm-3155 | $Bi_2O_3$ | 47.7 | − |
| 14 | 2dm-86 | $RhI_3$ | 48.4 | − |

Note: Red color represents the finally confirmed crystals with TFBs.

Table S2. 13 magnetic crystals with possible TFBs.

| No. | Mat-ID | Formula | Bandwidth (meV) | Model |
|---|---|---|---|---|
| 1 | 2dm-930 | $Al_2Te_3$ | 0.3 | − |
| 2 | 2dm-1371 | $MnF_3$ | 6.5 | Diatomic-Kagome |
| 3 | 2dm-5804 | $LiV_2F_9$ | 6.7 | − |
| 4 | 2dm-5206 | $Nb_3Cl_8$ | 14.1 | Breathing-Kagome |
| 5 | 2dm-1101 | $FeF_3$ | 16.3 | − |
| 6 | 2dm-5497 | $Nb_3I_8$ | 18.1 | − |
| 7 | 2dm-1128 | $NpBr_3$ | 22.8 | − |
| 8 | 2dm-3504 | $Na_2ZnCl_4O_3$ | 22.9 | Coloring-triangle |
| 9 | 2dm-390 | $MnBr_3$ | 26.7 | Diatomic-Kagome |
| 10 | 2dm-1976 | $VF_3$ | 27.9 | Honeycomb |
| 11 | 2dm-5839 | $Li_2Fe_3F_8$ | 43.1 | Kagome |
| 12 | 2dm-969 | $V_3F_8$ | 48.2 | Kagome |
| 13 | 2dm-1427 | $MnCl_3$ | 48.8 | − |

Note: $Na_2ZnCl_4O_3$ (labeled as blue) does not exist, but it inspires us to find the monolayer $III_2VI_3$ as coloring-triangle crystal.

Table S3. Monolayer $III_2VI_3$ with VI atoms at coloring-triangle lattice sites.

| VI \ III | B | Al | Ga | In | Tl |
|---|---|---|---|---|---|
| O | ✓ | ✓ | ✓ | ✓ | ✓ |
| S | ✓ | ✓ | ✓ | ✓ | ✗ |
| Se | ✓ | ✓ | ✓ | ✗ | ✗ |
| Te | ✓ | − | ✓ | − | ✗ |

Note, ✓: there is Kagome FB in monolayer $III_2VI_3$; ✗: there is not.
 − : the atomic structure does not exist in theory.

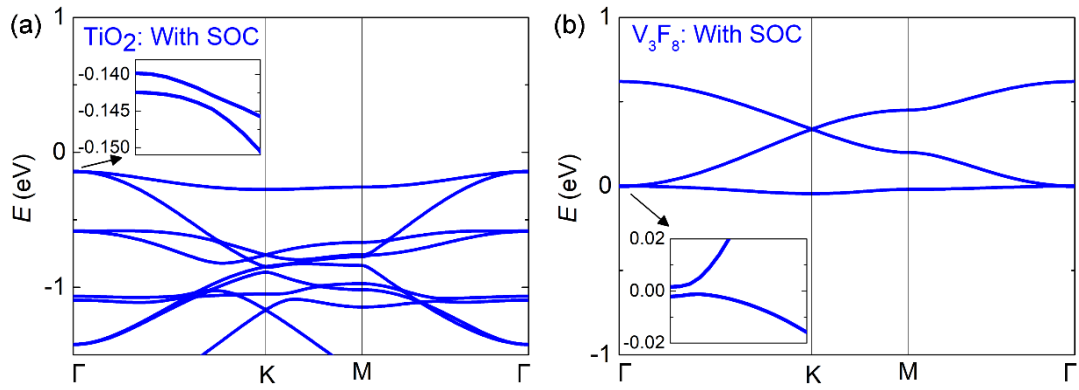

FIG. S1. Band structure with SOC of monolayer (a) TiO$_2$ and (b) V$_3$F$_8$.

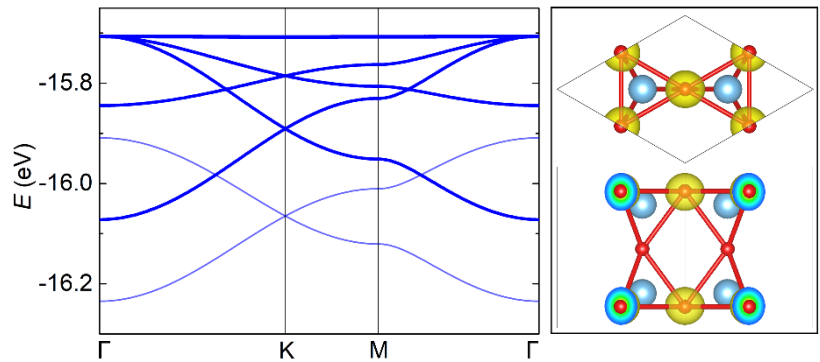

FIG. S2. Two sets of Kagome bands with degenerate TFBs in monolayer TiO$_2$ (left panel). The Kagome states are contributed by charge density from *s* orbitals of oxygen atoms (right panel).

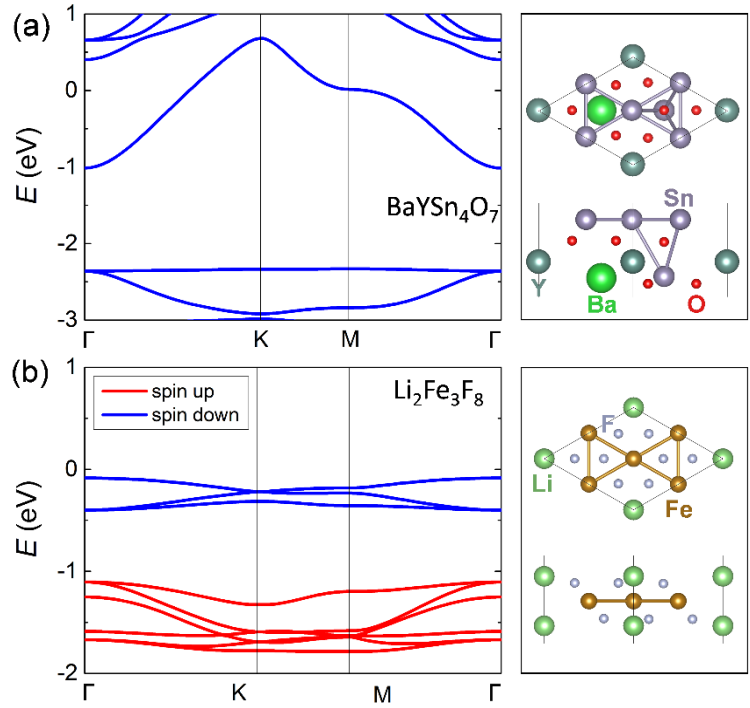

FIG. S3. Atomic and electronic structure of (a) nonmagnetic BaYSn$_4$O$_7$ with tin atoms on Kagome sites, and (b) ferromagnetic Li$_2$Fe$_3$F$_8$ with iron atoms on Kagome sites.

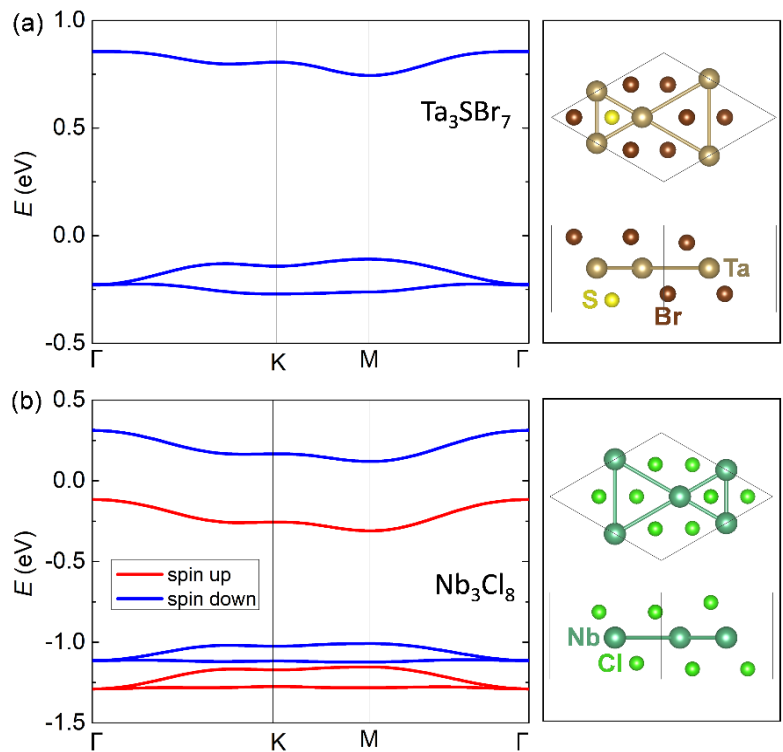

FIG. S4. Atomic and electronic structure of monolayer breathing-Kagome crystals (a) Ta$_3$SBr$_7$ and (b) Nb$_3$Cl$_8$.

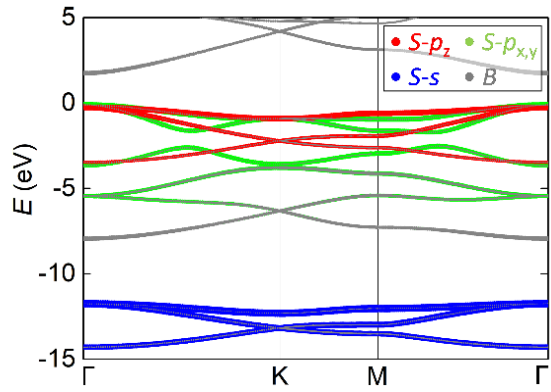

FIG. S5. Electronic structure and orbital composition of monolayer B$_2$S$_3$ without tensile strain.

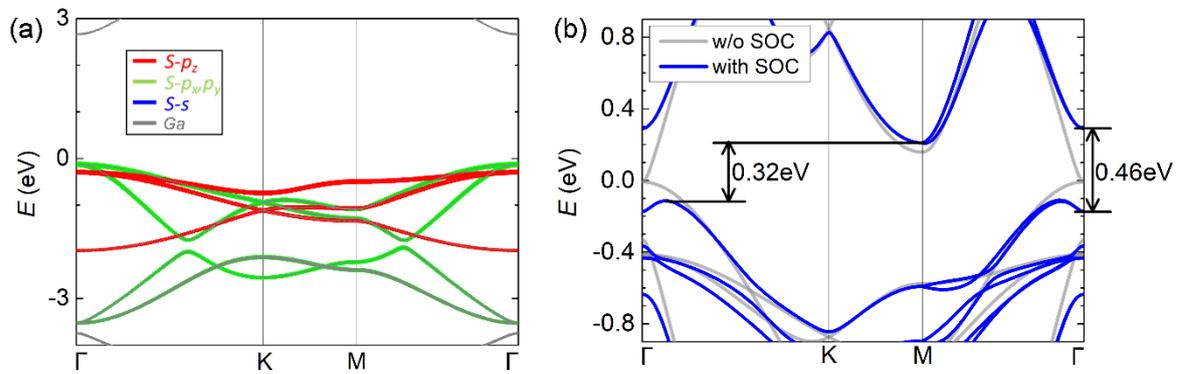

FIG. S6. (a) Kagome states from $p_z$ orbital of sulphur atoms in monolayer Ga$_2$S$_3$. (b) The large topological insulating gap induced by the strong SOC in monolayer Tl$_2$Te$_3$.

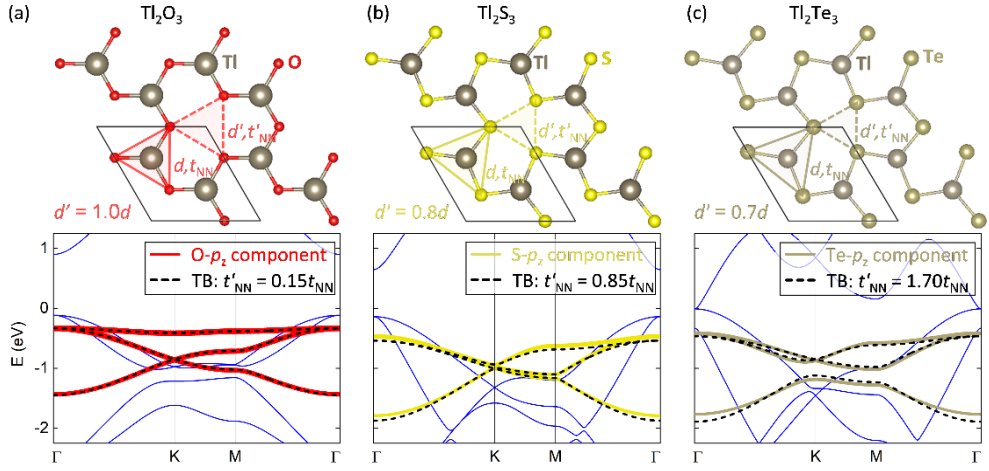

FIG. S7. Atomic and band structure of monolayer (a) Tl$_2$O$_3$, (b) Tl$_2$S$_3$, and (c) Tl$_2$Te$_3$. Upper panels: $d$ ($d'$) represents the distance between neighboring VI atoms bridged (not bridged) by a Tl atom, indicated by solid (dashed) lines. Lower panels: The VI-$p_z$ orbital weight of bands for Tl$_2$O$_3$, Tl$_2$S$_3$, and Tl$_2$Te$_3$, shown by the thickness of red, yellow, and tawny lines, respectively. Tight-binding coloring-triangle model with hopping integrals of $t'_{NN} = 0.15 t_{NN}$, $t'_{NN} = 0.85 t_{NN}$, and $t'_{NN} = 1.70 t_{NN}$ (black dashed lines) produces VI-$p_z$ bands of Tl$_2$O$_3$, Tl$_2$S$_3$, and Tl$_2$Te$_3$, respectively.

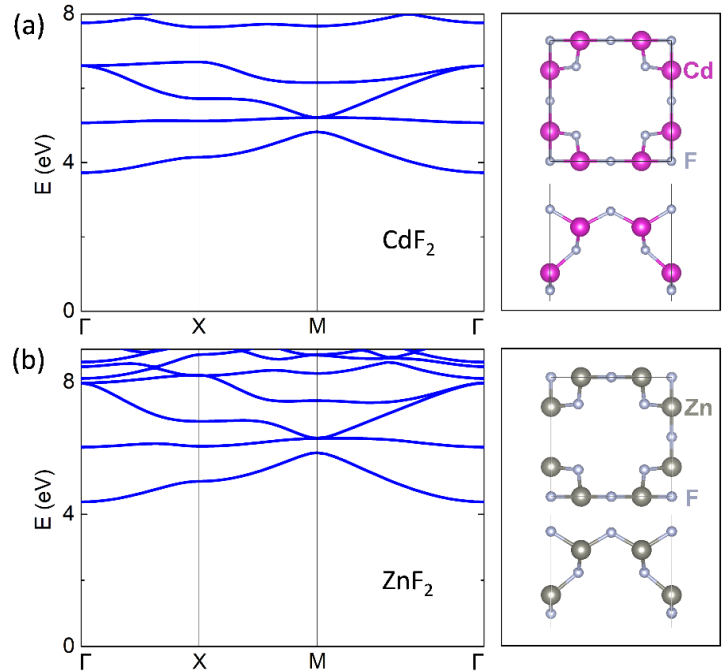

FIG. S8. Atomic and electronic structure of monolayer (a) CdF$_2$, and (b) ZnF$_2$.

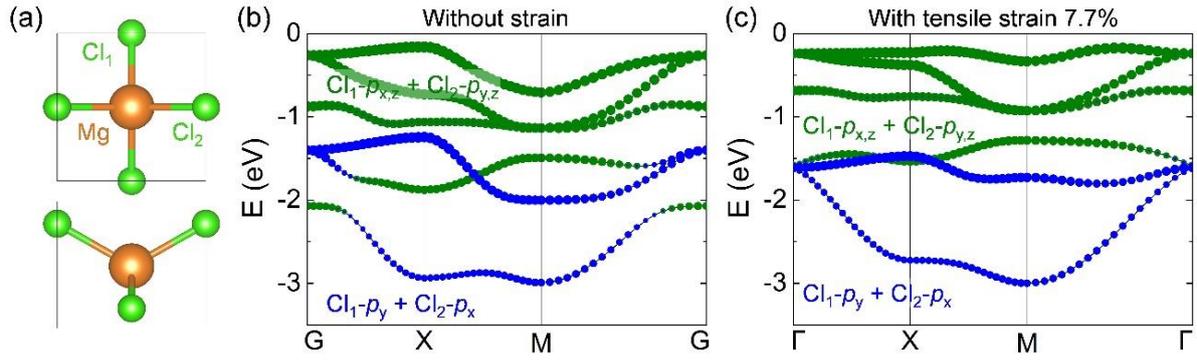

FIG. S9. Atomic and electronic structure of monolayer MgCl$_2$. (a) Orange and green balls represent magnesium and chlorine atoms, respectively. Black lines show the unit cell. (b) Band structure of MgCl$_2$ without strain. The blue and green dots represent the contribution from orbitals of (Cl$_1$-$p_y$, Cl$_2$-$p_x$) and (Cl$_1$-$p_{x,z}$, Cl$_2$-$p_{y,z}$), respectively. Fermi level is at zero energy. (c) Band structure of MgCl$_2$ with 7.7% tensile strain. The top four bands (green) are the same to that in diamond-octagon lattice model. The bottom two bands (blue) are from the checkerboard lattice model with $p$ orbitals, which is equivalent to the ($s$, $p$)-orbital square-lattice model with a diagonal $p$-orbital orientation [4].

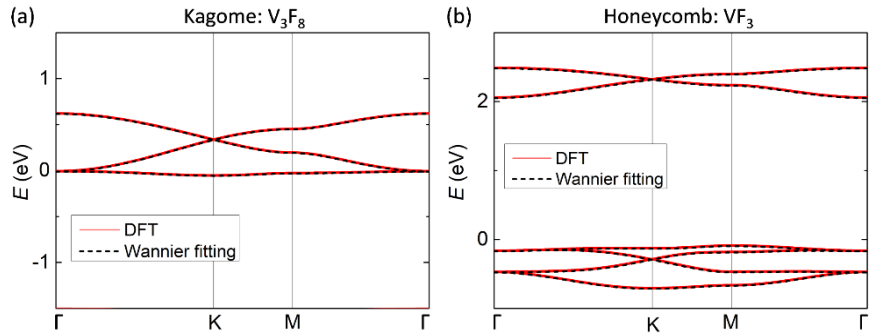

Fig. S10. Wannier construction (dashed lines) of *ab-initio* bands (solid lines) for monolayer (a) V$_3$F$_8$ and (b) VF$_3$ without SOC. The Wannier fitting for V$_3$F$_8$ (VF$_3$) used the orbitals of V-$d_{z2}$ (V-$d_{xy}$, $d_{x2-y2}$, $d_{z2}$), which have major contributions to the bands of concern.